\documentclass[12pt]{article}
\usepackage{array}
\begin{document}
  
\title {Bigravity in Kucha\u{r}'s Hamiltonian formalism. 2. The special case}
 \author {Vladimir O. Soloviev$^a$ and Margarita V. Tchichikina$^b$\\
 {\small \it $^a$Institute for High Energy Physics, 142 281, Protvino, Moscow region, Russia}\\
{\small \it e-mail: Vladimir.Soloviev@ihep.ru}\\
{\small \it $^b$Moscow State University, 119 899, Moscow, Russia}\\
{\small \it e-mail: chich@goa.bog.msu.ru}
}

\maketitle

\begin{abstract}
 It is proved, that, in order to avoid the ghost mode in  bigravity theory, it is sufficient to impose four conditions on the potential of interaction of the two metrics. First, the potential should allow its expression as a function of components of the two metrics' 3+1-decomposition. Second, the potential must satisfy the first order linear differential equations which are necessary for the presence of four first class constraints in  bigravity. Third, the potential should be a solution of the Monge-Amp\`ere equation, where the lapse and shift are considered as variables. Fourth, the potential must have a nondegenerate Hessian in the shift variables. The proof is based on the explicit derivation of the Hamiltonian constraints, the construction of Dirac brackets on the base of a part of these  constraints, and calculation of other constraints' algebra in these Dirac brackets. As a byproduct, we prove that these conditions are also sufficient in the massive gravity case. 
\end{abstract}

\section{Introduction}
Despite the fact that General Relativity (GR) for a very long time has stood as the Standard Model of gravity, a lot of effort has been undertaken to provide its generalizations, and  bigravity theory is one of them. N.~Rosen~\cite{Rosen} was probably the first who coined the term ``bimetric'' and who advocated a theory of gravitation with two spacetime metrics, but one of them was  nondynamical, i.e. the background one. Later, dynamics for the second metric was introduced~\cite{Salam}, and at first such a theory was called ``f-g gravity'' or ``strong gravity''.  The terms ``bigravity'' and ``multigravity'' appeared in works by T.~Damour, J.~Kogan, and their collaborators~\cite{Kogan}, where this theory was motivated  by the boom of extradimensional models from the theory side and  by the newly found effect of acceleration in the Universe expansion from the side of experiment. 
It was shown that in bigravity there were two gravitational fields, one of them was massless and the other was massive. Massive gravity is a longstanding challenge for theoreticians. Here S.~Deser remains an outstanding and permanent skeptic~\cite{BD}. In a recent article, Deser and Waldron~\cite{Deser} claim that even if a theory is ghostless, then it is  acausal. Certainly, in order to make final conclusions on the consistency of any theory with two or more spacetime metrics, it is necessary to provide a deep analysis of the causality problem, but here we do not plan to discuss it. Also, we are reminded~\cite{Deser} that this field of research has a rather long history, and remarkably, even the most fashionable variant of massive gravity~\cite{dRGT} has been proposed more than 40 years ago~\cite{WZ}. Nevertheless, we hope that there is some space for  bigravity.  Both  bigravity, and massive gravity seem to be able to explain the accelerated expansion of the Universe, and this is discussed in many papers, see, for example, Ref.~\cite{DM}.  

We think that the de Rham-Gabadadze-Tolley (dRGT) model~\cite{dRGT} of massive gravity and the Hassan-Rosen model of bigravity~\cite{HR} are very interesting and deserve a thorough study. In particular, it is necessary to obtain a more transparent analysis of their canonical structures, including deriving of all the constraints, calculating their algebra, classifying these constraints, and fixing the number of degrees of freedom. Obviously, a lot of articles with this work have already been published, we mention only a few~\cite{HR,HR093230,HR_bi,Kluson,HR2,Golovnev,Krasnov}, but the problem is difficult due to  the complexity and bulkiness of the proposed derivations coming from the nontrivial form of the potential with the matrix square root. So, these models are still under investigation (see especially the Conclusion for a discussion of the recent works).   

The goal of this article is to provide the Hamiltonian analysis of these new models in a way that is generalized as much as possible. For this purpose, we make three steps. First, we replace the dRGT potential by the most general one. Second, we replace massive gravity by  bigravity. Third, we replace ADM formalism~\cite{ADM} by the more general covariant Kucha\u{r}~\cite{Kuchar} approach.
 There are two kinds of matter present in formalism, each one interacts (the coupling is minimal) with its own metric, and our main results do not depend on the concrete form of the matter Lagrangians. We divided our work in two parts, which we call the general case and special case of bigravity. 

In the first part of this work~\cite{SolTch}, we obtain conditions which the potential of a general form $\tilde U(u,u^i,\eta_{ij},\gamma_{ij})$ should satisfy in order to have a theory with 4 first class constraints. The second class constraints are used for a construction of Dirac brackets. In those brackets, the algebra of first class constraints is the celebrated algebra of hypersurface deformations (see ADM~\cite{ADM} and Dirac~\cite{Dirac}). 

In this article, which is the second part of our work considered as a whole, we put new restrictions on the potential function in order to meet properties of the dRGT potential. Namely, in article ~\cite{SolTch} it was supposed that a matrix of second derivatives of the potential over variables $u,u^i$ was nondegenerate. Here we require that this matrix be degenerate and have rank 3, speaking more precisely, the matrix of second derivatives of the potential over variables $u^i$ is nondegenerate. At the same time, the potential is still  to fulfill conditions, found in Ref.~\cite{SolTch}, which are necessary for the existence of first class constraint algebra.

Really, there are at least three different algorithms which allow us to come to the results. The first one is to construct Dirac brackets on the base of the set of second class constraints. It is suitable that this set is not necessary to be complete. The second method is to use more Lagrangian multipliers, but work with the Poisson brackets. And the third way is to solve the second class constraints (may be implicitly) and express some variables as functions of the reduced set of canonical variables. We pay most attention to  the first method, i.e. in the main text we work with the Dirac brackets, but one can see in short how other methods work in the Appendixes.

The content of this article is as follows.

In Section 2 the notations and results of our previous article~\cite{SolTch} are reviewed. 

In Section 3 we outline in brief properties of the dRGT potential,
 interpret the dRGT conditions in our variables, and discuss consequences of these conditions. The $8\times 8$ matrix of the Poisson brackets among 8 constraints, considered in Ref.~\cite{SolTch} as nondegenerate, now becomes degenerate, consequences of this fact are analyzed, and the Dirac brackets are derived for the canonical variables (see Table~\ref{table1}). 

The results of Dirac bracket calculations among 6 constraints ${\cal R}$, ${\cal R}_i$, $\pi_u$, ${\cal S}$ are given in Section 4. 

In Section 5 the mathematical results on Monge-Amp\`ere equation provided in~\cite{Leznov} are applied, the crucial identity $\Theta^i=0$ is proved, and the Hassan-Rosen transformation is commented.

Section 6 contains discussion on quaternary constraint $\Psi$ and on Table~\ref{table} of Dirac brackets between the constraints. 
The number of gravitational degrees of freedom in bigravity for the potential with the given properties is proved to be 7.

In Section 7 we discuss the massive gravity case, i.e., a theory with only one dynamical metric, whereas the second one is a given solution of GR equations. In that case there are no constraints ${\cal R}$, ${\cal R}_i$ and the Hamiltonian is nonzero. 
The potential considered in this work gives 5 gravitational degrees of freedom for massive gravity.

The Conclusion contains a complete list of conditions imposed on the potential in this article, -- and  short comments on the related articles appeared after publication of the first part of this work~\cite{SolTch}.

We prefer in this article to use the same notations as in Ref.~\cite{SolTch}. In particular, for indices running from 0 to 3 we use small Greek letters, for indices running from 1 to 3 we use small Latin ones, spacetime metrics $f_{\mu\nu}$ and $g_{\mu\nu}$ have signature $-+++$. When the same letter is used for analogous quantities constructed with the first or with the second metric, then upper bar refers to the second one, i.e., 
$g_{\mu\nu}$. We significantly exploit here results obtained in Ref.~\cite{SolTch}.

\section{Kucha\u{r}'s approach in bigravity}
As it is in more detail presented in article~\cite{SolTch}, the bigravity action is a sum of two copies of the GR action, each given with its own matter as a source of gravity, minus a potential
\begin{equation}
{\cal L}={\cal L}^{(f)}+{\cal L}^{(g)}-\sqrt{-f}U( f_{\mu\nu}, g_{\mu\nu}).\label{eq:L_bi}
\end{equation}
\begin{equation}
{\cal L}^{(f)}=\frac{1}{16\pi G^{(f)}}\sqrt{-f}f^{\mu\nu}R_{\mu\nu}^{(f)}+{\cal L}_M^{(f)}(\psi^A, f_{\mu\nu}),\label{eq:Lf}\
\end{equation}
\begin{equation}
{\cal L}^{(g)}=\frac{1}{16\pi G^{(g)}}\sqrt{-g}g^{\mu\nu}R_{\mu\nu}^{(g)}+{\cal L}_M^{(g)}(\phi^A, g_{\mu\nu}),\label{eq:Lg}
\end{equation}
where $f$, $g$ are determinants of the first and second spacetime metrics, $R_{\mu\nu}^{(f)}$, $R_{\mu\nu}^{(g)}$ are their Ricci tensors, $G^{(f)}$, $G^{(g)}$ are the gravitational constants, ${\cal L}_M^{(f)}$, ${\cal L}_M^{(g)}$ are Lagrangians of the first and second matter, capital Latin letters are abstract indices for these matter fields.
 The potential 
\begin{equation}
\sqrt{-f}U( f_{\mu\nu},g_{\mu\nu}),
\end{equation}
having dimension $m^4$, is constructed as an ultralocal function of the two spacetime metrics.
Then the Hamiltonian of bigravity is also a sum of two copies of GR Hamiltonian plus the potential ultralocally depending on components of $3+1$ decomposition of the two metrics. In the Kucha\u{r} approach~\cite{Kuchar,Solo88}, where embedding variables $X^\mu=e^\mu(\tau,x^i)$ are involved in addition to metric, this decomposition is done with the help of basis $(n^\mu,e^\mu_i)$. Here $e^\mu_i=\partial X^\mu/\partial x^i$ are tangential vectors to the spatial hypersurface of state, and vector $n^\mu$ is a unit normal to the hypersurface (future directed). There are two metrics in bigravity, and so we are able to construct two future directed unit normals, and respectively two bases.  When decomposing a metric in its proper basis we get 6 independent components representing the induced metric on the given hypersurface ($\perp\perp$ and $\perp i$ components are -1 and 0), but when decomposing it in the other basis we get $6+4=10$ components. We take as a default the basis constructed with the help of first metric $f_{\mu\nu}$; then
\begin{eqnarray}
f^{\mu\nu}&=&-n^\mu n^\nu+\eta^{ij}e^\mu_i e^\nu_j,\\ 
g^{\mu\nu}&=&g^{\perp\perp}n^\mu n^\nu +g^{i\perp}(e^\mu_i n^\nu+e^\nu_i n^\mu)+\left(\gamma^{ij}+\frac{g^{\perp i}g^{\perp j}}{g^{\perp\perp}}\right)e^\mu_i e^\nu_j,
\end{eqnarray}  
and in the $g_{\mu\nu}$ basis denoted as $(\bar n^\mu, e^\mu_i)$, all formulas are changed:
\begin{eqnarray}
g^{\mu\nu}&=&-\bar n^\mu \bar n^\nu+\gamma^{ij}e^\mu_i e^\nu_j,\\ f^{\mu\nu}&=&f^{\perp\perp}\bar n^\mu \bar n^\nu +f^{i\perp}(e^\mu_i \bar n^\nu+e^\nu_i \bar n^\mu)+\left(\eta^{ij}+\frac{f^{\perp i}f^{\perp j}}{f^{\perp\perp}}\right)e^\mu_i e^\nu_j,
\end{eqnarray}  
and at the same time both induced metrics $\eta_{ij}=f_{\mu\nu}e^\mu_i e^\nu_j$, $\gamma_{ij}=g_{\mu\nu}e^\mu_i e^\nu_j$ stay unchanged. As usual, contravariant tensors are inverse matrices $\eta_{ij}\eta^{jk}=\delta^k_i$, $\gamma_{ij}\gamma^{jk}=\delta^k_i$. The matter fields  are  to be given by their components of $3+1$ decomposition too. Time is defined by parameter $\tau$, monotonically numerating hypersurfaces, and not by coordinate $X^0$,  contrary to the ADM notations. The components of this time vector field,
\begin{equation}
N^\mu=\frac{\partial X^\mu}{\partial\tau}=Nn^\mu+N^i e^\mu_i\equiv\bar N\bar n^\mu+\bar N^i e^\mu_i,
\end{equation}
are  lapse and shift (they are different for the two bases). 
In the ADM approach, where the embedding variables are not used, only one coordinate system $X^\mu$ is at work, and both metrics contain their
 lapses and shifts in the list of their components  in the coordinate basis: 
\begin{equation}
\bar N=\frac{1}{\sqrt{-g^{00}}},\quad \bar N_i=g_{0i},\quad N=\frac{1}{\sqrt{-f^{00}}},\quad N_i=f_{0i}.
\end{equation}
For the ADM case, we need 20 components, contrary to 16 for Kucha\u{r}'s. It is a disadvantage of the ADM approach in comparison to Kucha\u{r}'s one in bigravity case.To come back from Kucha\u{r}'s notations to ADM ones, we put
\begin{equation}
\tau=X^0, \qquad x^i=X^i,
\end{equation}
and then
\begin{equation}
N^\mu=\delta^\mu_0, \qquad e^\mu_i=\delta^\mu_i.
\end{equation}

Exploiting $(n^\mu,e^\mu_i)$ as a default basis to make calculations easier, we replace 4 components of the second metric by the following:
\begin{equation}
u=\frac{1}{\sqrt{-g^{\perp\perp}}},\qquad u^i=-\frac{g^{\perp i}}{g^{\perp\perp}},
\end{equation}
having the simple geometric meaning: $u$ is an inverse of a norm (calculated in the second metric) of vector $n^\alpha$, constructed as a unit normal (in the first metric) to the hypersurface, and $u^i$ are three projections (calculated in the second metric) of coordinate basis vectors onto this unit normal
\begin{equation}
u=\frac{1}{\sqrt{|g^{\mu\nu}n_\mu n_\nu|}}, \qquad u^i=\frac{g^{\mu\nu}n_\mu e_\nu^i}{\sqrt{|g^{\mu\nu}n_\mu n_\nu|}}.
\end{equation}
Then
\begin{equation}
\bar N=uN,\qquad \bar N^i=N^i+u^i N,\label{eq:uui}
\end{equation}
and in general case studied in Ref.~\cite{SolTch} the bigravity Hamiltonian takes a following form~\footnote{To be pedantic we should first introduce momenta conjugate to $N$ and $N^i$, and then the primary constraints saying these momenta are really zero. So, constraints ${\cal R}$, ${\cal R}_i$ arise as the secondary ones. The removal of these lapse and shift variables with their momenta from the list of canonical variables looks as a gauge, see Eq.(27) of article~\cite{SolTch}}
\begin{equation}
{\rm H}=\int d^3x\left(N{\cal R}+N^i{\cal R}_i \right).
\end{equation}
Here expressions
\begin{eqnarray}
{\cal R}&=&{\cal H}+u\bar{\cal H}+u^i\bar{\cal H}_i+\tilde U,\\
{\cal R}_i&=&{\cal H}_i+\bar{\cal H}_i,
\end{eqnarray}
are first class constraints satisfying in Dirac brackets the well-known hypersurface deformation algebra~\cite{Dirac}
\begin{equation}
 \{ {\cal R}(x),{\cal R}(y)\}_D=\left(\eta^{ik}(x){\cal R}_k(x)+ \eta^{ik}(y){\cal R}_k(y)\right)\delta_{,i}(x,y),%\label{eq:bialg1}
\end{equation}
\begin{equation}
  \{ {\cal R}_i(x),{\cal R}_k(y)\}_D={\cal R}_i(y)\delta_{,k}(x,y)+ {\cal R}_k(x)\delta_{,i}(x,y),%\label{eq:bialg2}
\end{equation}
\begin{equation}
  \{ {\cal R}_i(x),{\cal R}(y)\}_D={\cal R}(x)\delta_{,i}(x,y).%\label{eq:bialg3}
\end{equation}
The Dirac brackets were derived in article~\cite{SolTch} by inverting matrix of Poisson brackets for 8 second class constraints:
\begin{equation}
\pi_u=0,\quad \pi_{u^i}=0,\quad {\bar{\cal H}}+\frac{\partial\tilde{U}}{\partial u}=0,\quad {\bar{\cal H}}_i+\frac{\partial\tilde{U}}{\partial u^i}=0.%\label{eq:2сlass}
\end{equation}
The first quartet of them arises as primary constraints, and the second one as secondary. The lapse and shift functions, as in GR,  are Lagrangian multipliers standing before the first class constraints. In this work, as in ~\cite{SolTch}, we admit that the potential
 $\tilde{U}=\sqrt{\eta}U(u,u^i,\eta_{ij},\gamma_{ij})$ 
fulfills the following conditions
\begin{equation}
{ Q}^i_k\equiv 2\eta_{jk}\frac{\partial\tilde{U}}{\partial\eta_{ij}}+2\gamma_{jk}\frac{\partial\tilde{U}}{\partial\gamma_{ij}}-u^i\frac{\partial\tilde{U}}{\partial u^k}-\delta^i_k\tilde{U}=0,\label{eq:Q}
\end{equation}
\begin{equation}
{ Q}^\ell\equiv 2u^j\gamma_{jk}\frac{\partial\tilde{U}}{\partial\gamma_{k\ell}}-u^\ell u\frac{\partial\tilde{U}}{\partial u}+\left(\eta^{k\ell}-u^2\gamma^{k\ell}-u^k u^\ell \right)\frac{\partial\tilde{U}}{\partial u^k}=0,\label{eq:QQ}
\end{equation}
which are necessary for constraints ${\cal R}=0$, ${\cal R}_i=0$ to be first class (see Appendix A).

\section{dRGT-like potential in bigravity}

To simplify formulas of this article we introduce some new notations:
\begin{equation}
\tilde U''=\frac{\partial^2\tilde U}{\partial u^2}\quad\tilde U_i=\frac{\partial^2\tilde U}{\partial u\partial u^i}\quad {L}_{ij}= \frac{\partial^2 \tilde{U}}{\partial u^i\partial u^j},\quad \bar U^i=(L^{-1})^{ij}\tilde U_j,\label{eq:U}
\end{equation}
\begin{equation}
{\cal S}=\frac{\partial{\cal R}}{\partial u}=\bar{\cal H}+\frac{\partial\tilde U}{\partial u}, \qquad {\cal S}_i=\frac{\partial{\cal R}}{\partial u^i}=\bar{\cal H}_i+\frac{\partial\tilde U}{\partial u^i},\label{eq:S}
\end{equation}
\begin{equation}
K_{ij}(x,y)=
\{{\cal S}_i(x),{\cal S}_j(y)\},\qquad \bar K=L^{-1}KL^{-1}.
\end{equation}
It is claimed~\cite{HR} that secondary constraints ${\cal S}$, ${\cal S}_i$ cannot be solved for lapse and shift in the massive gravity or in the bigravity with the dRGT potential. In our approach instead of lapse $\bar N$ and shift $\bar N^i$ we use variables $u$, $u^i$, so this claim now sounds as
\begin{equation}
\frac{D({\cal S},{\cal S}_i)}{D(u,u^k)}=0. 
\end{equation}
It is evident from constraints structure (\ref{eq:S}) that this Jacobian is really a Hessian
\begin{equation}
\frac{D({\cal S},{\cal S}_i)}{D(u,u^k)}=
{\rm Det}\left(\frac{\partial^2\tilde U}{\partial u^a\partial u^b}\right)={\rm Det}\mathbf{L}=0, \label{eq:bfL}
\end{equation}
where we give a notation introduced in Eq.(94) of article~\cite{SolTch}. With the new notations (\ref{eq:U}) it is possible to write matrix $\mathbf{L}$ as follows
\begin{equation}
 \mathbf{L}=
\left(
\begin{array}{cc}
 \tilde U''
& \tilde U_j \\
\tilde U_i
& 
L_{ij}(x) 
\end{array}
\right).\nonumber
\end{equation}
Then the degeneracy of this matrix means that for the potential of dRGT-like form, which we call in this article as a potential of the special form, the following condition should be valid
\begin{equation}
 \tilde U_i\left({L}^{-1}\right)^{ij} \tilde U_j=\tilde U'',\label{eq:L}
\end{equation}
or
\begin{equation}
 \tilde U_i \bar U^i=\tilde U''.\label{eq:2-condition}
\end{equation}
The straightforward check of degeneracy of $\mathbf{L}$ and of nondegeneracy of $||L_{ij}||$ for the dRGT potential is  problematic. Our method is to suppose that they are valid and to consider their consequences. Let us remind that there are two other presumptions adopted in this article: 1) that dRGT potential can be written as a function of $3+1$-decomposition components of two metric tensors and 2) that this potential fulfills conditions (\ref{eq:Q}),  (\ref{eq:QQ}), which are necessary for existence of the first class constraint algebra of hypersurface deformations.

Introduced in our first article~\cite{SolTch} matrix $\mathbf{L}$ of Poisson brackets  between 8 constraints of the set  $\chi_A$ defined there by Eqs.(92)   becomes here degenerate because of Eq.(\ref{eq:2-condition}), and as a result cannot be used for the construction of  Dirac brackets~\cite{Dirac}. Two constraints from that set will not be included into the first set of second class constraints which will be denoted now as $\tilde \chi_A$,  $A=1,..,6$: 
\begin{equation}
\tilde\chi_A=\left(\pi_{u^i},{\cal S}_i\right),\label{eq:chi}
\end{equation}

Then the matrix of Poisson brackets for constraints $\tilde \chi_A$ will be 
\begin{equation}
||\{\tilde\chi_A(x),\tilde\chi_B(y)\}||=
\left(
\begin{array}{cc}
\mathbf{0} & -\mathbf{\tilde L}(x)\delta(x,y)\\
\mathbf{\tilde L}(x)\delta(x,y) & \mathbf{\tilde K}(x,y)
\end{array}\right),\nonumber
\end{equation}
where
\begin{equation}
 \mathbf{\tilde L}(x)=||L_{ij}(x)||,\nonumber
\end{equation}
\begin{equation}
\mathbf{\tilde K}(x,y)=
 ||K_{ij}(x,y)||.\nonumber
\end{equation}

As we suppose that matrix $\mathbf{\tilde L}$  is invertible than matrix  $||\{\tilde\chi_A,\tilde\chi_B\}||$ is invertible too, and its inverse has the following form 

\begin{equation}
\mathbf{\tilde C}(x,y)
=\left(
\begin{array}{cc}
 \mathbf{\tilde L}^{-1}(x)\mathbf{\tilde K}(x,y)\mathbf{\tilde L}^{-1}(y) & \mathbf{\tilde L}^{-1}(x)\delta(x,y)  \\
 -\mathbf{\tilde L}^{-1}(x)\delta(x,y) & \mathbf{0}
\end{array}
\right).\nonumber
\end{equation}
Dirac brackets are now defined as follows
\begin{equation}
 \{F,G\}_{\tilde D}=\{F,G\}-\int dx\int dy\{F,\tilde\chi_A(x)\}\mathbf{\tilde C}^{AB}(x,y)\{\tilde\chi_B(y),G\},\label{eq:DB}
\end{equation}
where tilde is inserted to mark a difference from another Dirac brackets used in~\cite{SolTch}. In more detail,
\begin{eqnarray}
 \{F,G\}_{\tilde D}&=&\{F,G\}-\int dx\int dy\{F,\pi_{u^i}(x)\}\bar K^{ij}(x,y)\{\pi_{u^j}(y),G\}\nonumber\\
&-&\int dx\{F,\pi_{u^i}(x)\}({L}^{-1})^{ij}(x)\{{\cal S}_{j}(x),G\}\nonumber\\
&+&\int dx\{F,{\cal S}_{i}(x)\}({L}^{-1})^{ij}(x)\{\pi_{u^j}(x),G\}.
\end{eqnarray}

In case both functionals $F$, $G$ do not depend on $u^i$, these brackets coincide with Poisson ones. As second class constraints 
\begin{equation}
{\cal S}_i=0
\end{equation}
can be explicitly solved for $u^i$,  we can consider these
variables as functions of variables $u$, $\eta_{ij}$, $\gamma_{ij}$ and of $\bar{\cal H}_i$, see Appendix~\ref{solv_const}. Functions $N$, $N^i$ are Lagrangian multipliers for constraints  ${\cal R}=0$, ${\cal R}_i=0$ and have zero Dirac brackets with all functionals. Primary constraint $\pi_u=0$ is to be included into the Hamiltonian with a corresponding Lagrangian multiplier 
\begin{equation}
{\rm H}=\int d^3x \left(N{\cal R}+N^i{\cal R}_i+\lambda\pi_u \right),
\end{equation}
whereas secondary constraint  ${\cal S}=0$ reappears from the requirement of the primary constraint conservation:
\begin{equation}
\dot \pi_u=\{\pi_u,{\rm H}\}_{\tilde D}=-{\cal S}=0.
\end{equation}  
Here we stand with 13 pairs of canonically conjugate gravitational variables $(u,\pi_u)$, $(\eta_{ij},\Pi^{ij})$, $(\gamma_{ij},\pi^{ij})$, and with some pairs of matter variables, $(\psi_A,\Pi^A)$, $(\phi_A,\pi^A)$,  of course. There is a set of constraints which are weakly zero (i.e. they could not be treated as zero before calculation of Dirac brackets): 
\begin{equation}
\pi_u\approx 0,\quad {\cal S}\approx 0,\quad {\cal R}\approx 0,\quad {\cal R}_i\approx 0.\label{eq:para}
\end{equation}
In the bigravity we expect that the last four constraints will stay first class and they will provide the diagonal diffeomorphism invariance. The first constraint is the most trivial and serves to exclude $u,\pi_u$ degree of freedom, but to fulfill this task we are to find a companion constraint. The second constraint ${\cal S}$ may be called as the Hamiltonian constraint because it is quadratic in momenta, however ${\cal R}$ shares this property.
Next step is to find the full set of constraints and to separate them into first and second class sets. To solve this problem we should calculate Dirac brackets between the constraints.

\begin{table}
\setlength{\extrarowheight}{5pt}
\begin{tabular}{|c||c|c|c|c|c|c|c|}
\hline 
$\{,\}_{\tilde D}$  & 
$u(y)$ & 
$\pi_u(y)$ & 
$u^k(y)$ & 
$\eta_{mn}(y)$ & 
$\Pi^{mn}(y)$ & 
$\gamma_{mn}(y)$ & 
$\pi^{mn}(y)$ \\ 
\hline \hline
$u(x)$ & 0 & $1$ & 0 &  0 & 0 & 0 & 0\\
\hline
$\pi_u(x)$ &  $-1$ & 0 & $\bar U^i$ &  0 & 0 & $0$ & 0 \\
\hline
$u^i(x)$  & 0 & $-\bar U^i$ & $\bar K^{ik}$ & 0 & $A^{imn}$ & $B^i_{mn}$ & $C^{imn}$  \\
\hline
$\eta_{ij}(x)$ & 0 & 0 & 0 & 0 & $\delta_{ij}^{mn}$& 0 & 0 \\ 
\hline
$\Pi^{ij}(x)$ 
& 0 & 0 & $-A^{kij}$ &  $-\delta_{mn}^{ij}$ &  0 &  0 & 0 \\
\hline
$\gamma_{ij}(x)$ &  0 & 0 & $-B^k_{ij}$ & 0 & 0 & 0 & $\delta_{ij}^{mn}$ \\
\hline
$\pi^{ij}(x)$ & 0 & 0 & $-C^{kij}$ &  0 & 0 & $-\delta^{ij}_{mn}$ & 0  
\\
\hline
\end{tabular} 
\caption{Dirac brackets for main variables}\label{table1}
\end{table}

We summarize formulas for Dirac brackets between the canonical variables in the table, see Table~\ref{table1}.  Some notations are shown below:
 \begin{eqnarray}
A^{imn}
&=&
-(L^{-1})^{ij}
\frac{\partial^2\tilde U}{\partial u^j\partial\eta_{mn}}
\delta(x,y),\\
B^i_{mn}
&=&
-(L^{-1})^{ij}(x)
\{\bar{\cal H}_j(x),\gamma_{mn}(y)\},\\
C^{imn}&=&-(L^{-1})^{ij}\left(\frac{\partial^2\tilde U}{\partial u^j\partial\gamma_{mn}}+
\{\bar{\cal H}_j(x),\pi^{mn}(y)\}\right).
\end{eqnarray}

\section{Dirac brackets for secondary constraints}
In Appendix A we display the Poisson brackets between constraints ${\cal R}$ and ${\cal R}_i$. After replacing these Poisson brackets by Dirac ones we can treat constraints 
\begin{equation}
\pi_{u^i}=0,\qquad {\cal S}_i=0,
\end{equation}
as strongly equal to zero, i.e. we can omit them and obtain the following
\begin{equation}
 \{ {\cal R}(x),{\cal R}(y)\}_{\tilde D}=\left[(\eta^{ik}{\cal R}_k+uu^i{\cal S})(x)+ (\eta^{ik}{\cal R}_k+uu^i{\cal S})(y)\right]\delta_{,i}(x,y),\label{eq:alg1}
\end{equation}
\begin{equation}
  \{ {\cal R}_i(x),{\cal R}_k(y)\}_{\tilde D}={\cal R}_i(y)\delta_{,k}(x,y)+ {\cal R}_k(x)\delta_{,i}(x,y),\label{eq:alg2}
\end{equation}
\begin{equation}
  \{ {\cal R}_i(x),{\cal R}(y)\}_{\tilde D}={\cal R}(x)\delta_{,i}(x,y)+u_{,i}{\cal S}\delta(x,y).\label{eq:alg3}
\end{equation}

For the general potential the Dirac bracket of constraint ${\cal S}$  with itself is nonzero:
\begin{eqnarray}
\{
{\cal S}(x),{\cal S}(y)
\}_{\tilde D}
&=&
\{
{\cal S}(x),{\cal S}(y)
\} 
+\bar U^i(x)\{{\cal S}_i(x),{\cal S}_j(y)\}\bar U^j(y)\nonumber\\
&-&\{{\cal S}(x),{\cal S}_i(y)\}\bar U^i(y)+\{{\cal S}(y),{\cal S}_i(x)\}\bar U^i(x)\nonumber\\
&=& (\Theta^i-\bar U^i{\cal S})(x)\delta_{,i}(x,y)-(\Theta^i-\bar U^i{\cal S})(y)\delta_{,i}(y,x),
\end{eqnarray}
where
\begin{equation}
\Theta^i =
\left(\bar U^k\hat{D}\left(\delta^i_k-2\gamma_{jk}\frac{\partial}{\partial\gamma_{ij}} \right)-\gamma^{ij}\frac{\partial}{\partial u^j}\right)\tilde U,\label{eq:Theta_definition}
\end{equation}
and a new notation has been introduced
\begin{equation}
\hat{D}=\frac{\partial}{\partial u}-\bar U^i\frac{\partial}{\partial u^i}.
\end{equation}
Evidently, if the following condition is valid for some quantity $Z$, 
\begin{equation}
\hat{D} Z=0,
\end{equation}
then this quantity $Z$ does not depend on $u$ when constraints ${\cal S}_i=0$ are taken into account.

The Dirac bracket between constraints ${\cal S}$ and ${\cal R}$ is, in general,  not weakly equal to zero:
$$
\{{\cal R}(x),{\cal S}(y)\}_{\tilde D}=\left(u^i- u\bar U^i\right){\cal S}(x)\delta_{,i}(x,y)-\left( u(\bar U^i{\cal S})_{,i}+\Omega\right)\delta(x,y),
$$
and so, provides us with a new (tertiary) constraint $\Omega=0$ which is called as secondary in Refs.~\cite{HR,HR2}:
\begin{eqnarray}
\Omega
&=& 
  \left( u\hat{D}-1\right)[\tilde U,\bar{\cal H}] +\hat{D}[\tilde U,{\cal H}]\nonumber\\
&+& \left((\bar U_{k|l}+\bar U_{l|k})+(u_{k|l}+u_{l|k})\hat{D}\right)\frac{\partial\tilde U}{\partial\gamma_{kl}}
\nonumber\\
&+&  u_{,i}\bar U^i\frac{\partial\tilde U}{\partial u}-\left(u^k\bar U^i_{|k}-\bar U^k u^i_{|k}+\gamma^{ik} u_{,k} \right)\frac{\partial\tilde U}{\partial u^i}\nonumber\\
&-& \left(
u^i\hat{D}\tilde U
+ u\gamma^{ik}\frac{\partial\tilde U}{\partial u^k}
+2\bar U^j\gamma_{jk}\frac{\partial\tilde U}{\partial\gamma_{ik}} 
\right)_{,i},\label{eq:Omega}
\end{eqnarray}
here square brackets denote coefficients standing before $\delta$-function in  the related ultralocal Poisson brackets:
$$
\{\tilde U(x),\bar{\cal H}(y)\}=[\tilde U,\bar{\cal H}](x)\delta(x,y),\qquad \{\tilde U(x),{\cal H}(y)\}=[\tilde U,{\cal H}](x)\delta(x,y),
$$
and indices of quantities $u^i$, $\bar U^i$ are moved down by means of metric $\gamma_{ij}$.
First class constraints ${\cal R}_i$, as usual, generate spatial diffeomorphisms 
\begin{eqnarray}
\{
{\cal S}(x),{\cal R}_i(y)
\}_{\tilde D}&=&-{\cal S}(y)\delta_{,i}(y,x),\nonumber\\
\{
\Omega(x),{\cal R}_i(y)
\}_{\tilde D}&=&-{\Omega}(y)\delta_{,i}(y,x)\label{eq:3diff}
\end{eqnarray}
It is possible also to calculate the following brackets
\begin{eqnarray}
\{{\cal R}(x),\pi_u(y)\}_{\tilde D}&=&{\cal S}(x)\delta(x,y),\nonumber\\
\{\Omega(x),\pi_u(y)\}_{\tilde D}&=& \Theta^i(x)\delta_{,i}(x,y)-\Theta^i(y)\delta_{,i}(y,x). \label{eq:ThetaOmega}
\end{eqnarray}

In  Ref.~\cite{HR2} it was proved that ${\cal S}$ commutes with itself for the dRGT potential. So, it is reasonable to hope that our potential should satisfy condition $\Theta^i=0$, i.e.
\begin{equation} 
\left(\bar U^k\hat{D}\left(\delta^i_k-2\gamma_{jk}\frac{\partial}{\partial\gamma_{ij}} \right)-\gamma^{ij}\frac{\partial}{\partial u^j}\right)\tilde U=0.\label{eq:Theta}
\end{equation}
 The two requirements might be fulfilled at once by this: 
\begin{enumerate}
\item the tertiary (called as secondary in articles~\cite{HR,HR093230,HR_bi,HR2}) constraint will appear;
\item this appearing constraint will not depend on variable $u$.
\end{enumerate}
Eq.~(\ref{eq:Theta}) will be proved in the next Section.
If $\Theta^i$ were nonzero, then $\Omega=0$ would be an equation fixing variable $u$ and there would be 3 second class and 4 first class constraints for 13 pairs of gravitational canonical variables in bigravity. This would give  $7+1/2$ gravitational degrees of freedom for the bigravity, and $5+1/2$ for the massive gravity in line with the conclusion  first made by Kluson~\cite{Kluson}.

\section{Transformation of variables}
Previous sections were written before we discovered Comelli et al articles~\cite{Comelli2013,Comelli2012}. Their analysis refers  to the case of the massive gravity, the bigravity is not considered. The main point is to use the results by Fairlie and Leznov who found the implicit exact solution of the Monge-Amp\`ere equation~\cite{Leznov}.  First, we can see that $(1,-\bar U^i)$ is a null-vector of the Hessian matrix 
$$
\mathbf{L}=\left|\left|\frac{\partial^2 \tilde U}{\partial u^a\partial u^b}\right|\right|,
$$
so, in order to be close to article~\cite{Leznov}, we introduce a new notation:
$$
\xi^i=-\bar U^i.
$$
It is suitable to replace potential $\tilde U$ by function $V(\xi^i,\eta_{ij},\gamma_{ij})$, such that
\begin{equation}
\frac{\partial\tilde U}{\partial u^i}=\frac{\partial V}{\partial \xi^i},\qquad \frac{\partial\tilde U}{\partial u}=V-\xi^i\frac{\partial V}{\partial \xi^i},\label{eq:dU}
\end{equation}
then it follows that
\begin{equation}
\hat D\tilde U= \left(\frac{\partial}{\partial u}-\bar U^k\frac{\partial}{\partial u^k}\right)\tilde U=\left(\frac{\partial}{\partial u}+\xi^k\frac{\partial}{\partial u^k}\right)\tilde U=V.\label{eq:VDU}
\end{equation}
Integrability condition for Monge-Amp\`ere equation
\begin{equation}
{\rm Det}\left|\left|\frac{\partial^2\tilde U}{\partial u^a\partial u^b}\right|\right|=0\label{eq:integrability}
\end{equation}
is as follows (see~\cite{Leznov} for details) 
\begin{equation}
\hat D\xi^i=\frac{\partial \xi^i}{\partial u}+\xi^k\frac{\partial \xi^i}{\partial u^k}=0.\label{eq:Dxi}
\end{equation}
At last, it is proved in Ref.~\cite{Leznov} that any implicit solution for system of equations (\ref{eq:integrability})  can be written as follows~\cite{Leznov}
\begin{equation}
u^i-u\xi^i=(V^{-1})^{ik}\frac{\partial W}{\partial\xi^k},\label{eq:implicit}
\end{equation}
where $W=W(\xi^i,\eta_{ij},\gamma_{ij})$ is an arbitrary function, and
$$ 
V_{ik}=\frac{\partial^2 V}{\partial \xi^i\partial \xi^k}.
$$
If we apply differential operator $\hat D$ to Eq.~(\ref{eq:QQ}), and take into account (\ref{eq:dU}) and  (\ref{eq:VDU}) we get the following 
\begin{equation}
2\xi^j\gamma_{jk}\frac{\partial\tilde{U}}{\partial\gamma_{k\ell}}
+2u^j\gamma_{jk}\hat D\frac{\partial \tilde U}{\partial\gamma_{k\ell}}
-(\xi^\ell u+u^\ell)\left(V-\xi^i\frac{\partial V}{\partial \xi^i}\right)
-\left(2u\gamma^{k\ell}+\xi^k u^\ell+u^k \xi^\ell \right)\frac{\partial{V}}{\partial \xi^k}=0,
\end{equation}
then after repeating this operation and dividing the result by factor 2 we obtain
\begin{equation}
2\xi^j\gamma_{jk}\hat D\frac{\partial \tilde U}{\partial\gamma_{k\ell}}
+u^j\gamma_{jk}(\hat D)^2\frac{\partial \tilde U}{\partial\gamma_{k\ell}}
-\xi^\ell V-\gamma^{k\ell}\frac{\partial{V}}{\partial \xi^k}=0.
\end{equation}
Now  
\begin{equation}
\hat D\frac{\partial \tilde U}{\partial\gamma_{k\ell}}=\frac{\partial V}{\partial\gamma_{k\ell}}-\frac{\partial\xi^m}{\partial\gamma_{k\ell}}\frac{\partial V}{\partial\xi^m},
\end{equation}
 and as this expression depends on $u,u^i$ only through function $\xi^i(u,u^i,\gamma_{ij},\eta_{ij})$, and given Eq.~(\ref{eq:Dxi}), it is annihilated by operator $\hat D$, so  Eq.~(\ref{eq:Theta}) is proved.

It is interesting to compare the results given above  with the argumentation given in  pioneering work~\cite{HR093230}. Variables $u^i$ here play the role analogous to $N^i$ in Ref.~\cite{HR093230}, $u$ is similar to $N$, and variables $\xi^i$ or $\bar U^i$  are analogous to $n^i$. Two of the three criteria for the existence of the Hamiltonian constraint ${\cal S}$ formulated in Ref.~\cite{HR093230} are fulfilled here as one can see from Eqs. (\ref{eq:dU}) and (\ref{eq:implicit}). It is proved in Ref.~\cite{HR093230} that their third criterion is fulfilled automatically when the other two are satisfied. And let us consider a function
$$
\frac{\partial\tilde U}{\partial u}=V-\xi^i\frac{\partial V}{\partial \xi^i}
$$
on the surface of constraints ${\cal S}_i$, then
$$
{\cal V}=\left.\frac{\partial\tilde U}{\partial u}\right|_{{\cal S}_i=0}=V-\xi^i\bar{\cal H}_i, \qquad \frac{\partial{\cal V}}{\partial\xi^i}\approx 0,
$$
therefore on this constraint surface we have
$$
\frac{\partial{\cal S}}{\partial\xi^i}\approx 0,
$$
as it has been stated in Ref.~\cite{HR093230}.

\section{Dirac brackets for tertiary and quaternary constraints}

\begin{table}
\setlength{\extrarowheight}{5pt}
\begin{tabular}{|c||c|c|c|c|c|c|}
\hline 
$\{,\}_{\tilde D}$  & $\pi_u(y)$ & $\Psi(y)$& $\Omega(y)$ & ${\cal S}(y)$ & ${\cal R}(y)$ & ${\cal R}_j(y)$ \\ 
\hline \hline
$\pi_u(x)$ &  0 &$\ne 0$& $-\hat\Theta=0$
& $0$ & $\approx 0$
& $0$ \\
\hline
$\Psi(x)$  &$\ne 0$ & & && & \\
\hline
$\Omega(x)$ & $\hat\Theta=0$& 
&  & $\ne 0$ & $\Psi\approx 0$ & $\approx 0$ \\
\hline
${\cal S}(x)$ 
& 0 & & $\ne 0$ & $\hat\Theta=0$
 & 
$\approx 0$
& $\approx 0$
\\
\hline
${\cal R}(x)$ & $\approx 0$&
& $-\Psi\approx 0$ & 
$\approx 0$
& 
$\approx 0$ & 
$\approx 0$\\
\hline
${\cal R}_i(x)$ & 0&  & $\approx 0$ & 
$\approx 0$ & 
$\approx 0$
& 
$\approx 0$
\\
\hline
\end{tabular} 
\caption{Dirac brackets for constraints}\label{table}
\end{table}

For the potential satisfying  $\Theta^i=0$ we are to continue the process of calculating Dirac brackets between constraints in looking for new constraints.  
Unfortunately, expressions for constraints evidently become more complicated with each step. We derived tertiary constraint $\Omega$ above, see Eq. (\ref{eq:Omega}). In order to preserve this constraint in the process of evolution we are to ensure the following condition
\begin{equation}
\dot\Omega=\{\Omega, {\rm H}\}_{\tilde D}=\int N\{\Omega,{\cal R}\}_{\tilde D}+N^i\{\Omega,{\cal R}_i\}_{\tilde D}+\lambda\{\Omega,\pi_u\}_{\tilde D}\approx 0.\label{eq:difur}
\end{equation}
As the second and the third Dirac brackets here are zero on the surface of already known constraints, due to Eqs.~(\ref{eq:3diff}), (\ref{eq:ThetaOmega}),  it is necessary to calculate only the first bracket:
\begin{equation}
\{\Omega(x),{\cal R}(y)\}_{\tilde D}=\Psi(x)\delta(x,y),
\end{equation}
that will give us a new (quaternary) constraint
\begin{equation}
\Psi\approx 0.
\end{equation}
The Jacobi identity 
$$
 \{
\Omega,
\{\pi_u,{\cal R}\}_{\tilde D}
\}_{\tilde D}
+\{
{\cal R},\{\Omega,\pi_u\}_{\tilde D}\}_{\tilde D}
+\{\pi_u,
\{{\cal R},\Omega\}_{\tilde D}\}_{\tilde D}=0,
$$
if we take into account Eqs.~(\ref{eq:ThetaOmega}), gives us the following result 
\begin{equation}
\{
\pi_u(y),
\Psi(x)
\}_{\tilde D}\delta(x,z)= -\{
\Omega(x),{\cal S}(z)
\}_{\tilde D}\delta(z,y),\label{eq:OS}
\end{equation}
 so it is evident that if $\{\Omega,{\cal S}\}_{\tilde D}$ is nonzero, then $\{\pi_u,\Psi\}_{\tilde D}$ is nonzero also. 
But it is easy to check that there are nonzero terms in $\{\Omega,{\cal S}\}_{\tilde D}$, for example,
\begin{equation}
[[V,{\cal H}],\bar{\cal H}]=\frac{4\kappa^{(f)}\kappa^{(g)}}{\sqrt{\eta\gamma}}\left(\Pi_{ij}-\eta_{ij}\frac{\Pi}{2}\right)\frac{\partial^2 V}{\partial\eta_{ij}\partial\gamma_{mn}} \left(\pi_{mn}-\gamma_{mn}\frac{\pi}{2}\right)\ne 0,\label{eq:VHH}
\end{equation}
that cannot be canceled by other terms, and does not appear in other constraints. Therefore both Dirac brackets in Eq.(\ref{eq:OS}) are nonzero. Of course, in looking for the so-called partially massless case, see Refs.~\cite{Deser2,RHRT,PMBI1,PMBI2,PMBI3}, one needs a more detailed study.

There is a potential problem  if $\Psi(x)$ occurs not a function, but a differential operator, even after taking into account all constraints. Then Eq.~(\ref{eq:difur})  can be solved for $u$, though the solution will depend on $N$.

We combine schematically all the results of our calculations of the Dirac brackets in the table (see Table~\ref{table}),  these results are displayed in the text above in more detail. One can see that there are 3 blocks standing on the diagonal of the table: two nondegenerate $2\times 2$ matrices for two pairs of the second class constraints $(\pi_u,\Psi)$ serving to exclude $u,\pi_u$ degree of freedom, $(\Omega,{\cal S})$ serving to exclude the ghost degree of freedom and one degenerate $4\times 4$ matrix for the first class constraints $({\cal R},{\cal R}_i)$ providing the diagonal diffeomorphism invariance.  Therefore, for 13 pairs of the gravitational variables we have 4 first class constraints and 4 second class constraints,  this corresponds to 7 degrees of freedom for the bigravity.

\section{Hamiltonian approach to massive gravity}
Let us consider now the case when the first metric is a fixed GR solution (in general, with  matter sources, and maybe, with a cosmological constant) and only the second one  is dynamical. 
We call such theories as the massive gravity or the bimetric gravity following N.~Rosen~\cite{Rosen}.  Recently they have got a revival~\cite{Rubak,Blas,MMMM,Hinter}.  
Given metric $f_{\mu\nu}$ and embedding variables $e^\alpha(t,x^k)$ we are to treat  $N(t,x^k)$, $N^i(t,x^k)$, $\eta_{ij}(t,x^k)$ as the known functions. For the starting Hamiltonian we use the following expression
\begin{equation}
{\rm H}=\int d^3x \left(N\left(u{\bar{\cal H}}+u^i{\bar{\cal H}}_i+\tilde{U} \right)
+ N^i\bar{\cal H}_i +\lambda\pi_u+\lambda^i\pi_{u^i}\right).
\end{equation}
It is similar to bigravity Hamiltonian given that ${\cal H},{\cal H}_i$ are now zero, as metric $f_{\mu\nu}$ is now a fixed solution of GR equations. Of course, the Poisson brackets now do not involve variational derivatives over $\eta_{ij}$. There are 4 primary constraints
\begin{equation}
\pi_{u}=0, \quad \pi_{u^i}=0.\label{eq:piu}
\end{equation}
and, in their turn, they generate 4 secondary constraints
\begin{equation}
{\bar{\cal H}}+\frac{\partial\tilde{U}}{\partial u}\equiv {\cal S}=0,\qquad {\bar{\cal H}}_i+\frac{\partial\tilde{U}}{\partial u^i}\equiv {\cal S}_i=0,\label{eq:last}
\end{equation}
where $\tilde{U}=\tilde U(u, u^i, \gamma_{ij}, \eta_{ij})$.
The Poisson brackets matrix of these constraints should be degenerate for the dRGT potential as in the bigravity case. Then we again construct the Dirac brackets (of course, they are a bit more compact than in the bigravity, as the Poisson brackets now do not involve variational derivatives over $\frac{\delta}{\delta\eta_{ij}}$) by inverting  $6\times 6$ matrix constructed of the second class constraints  
$\tilde \chi_A=0$ (see Eqs.~(\ref{eq:chi})). As the Hamiltonian is a first class quantity we can drop out the second class constraints to simplify its form
 \begin{equation}
{\rm H_{reduced}}=\int d^3x\left( N\left( u\bar{\cal H}+\tilde U \right)-\bar N^i\frac{\partial \tilde U}{\partial u^i}+\lambda\pi_u\right).
\end{equation}
Primary constraint $\pi_u$ is still present in the reduced Hamiltonian accompanied by its Lagrangian multiplier $\lambda$, and secondary constraint ${\cal S}$ reappears as a condition of preservation of $\pi_u$ in evolution  
\begin{equation}
\dot \pi_u=\{
\pi_u,{\rm H_{reduced}}
\}_{\tilde D}=
-N{\cal S}
=0.
\end{equation}
Constraint ${\cal S}$ can be solved for the ghost variable. For consistency of the evolution we should calculate the Dirac bracket of the secondary constraint with the Hamiltonian and put it equal to zero.
Applying Eq.~(\ref{eq:DB}) we can write this Dirac bracket in terms of the Poisson ones
\begin{eqnarray}
\{ 
{\cal S}(x),
{\rm H_{reduced}}
\}_{\tilde D}
&=&
\{
 {\cal S}(x),
{\rm H_{reduced}} 
\}
-\bar U^i(x)
\{
{\cal S}_i(x),
{\rm H_{reduced}}
\}\nonumber\\
&-&\int d^3y\left(\bar U^i(x)\{ {\cal S}_i(x),{\cal S}_j(y)\}+\{{\cal S}(x),{\cal S}_j(y)\} \right)\bar N^j(y).\nonumber
\end{eqnarray}
The result is a new expression
\begin{equation}
 \{ {\cal S}(x),{\rm H_{reduced}} \}_{\tilde D}=\tilde\Omega,
\end{equation}
which is necessary to be zero
\begin{eqnarray}
 \tilde\Omega&=&
N\left(u\hat{D}-1\right)[\tilde U,\bar{\cal H}] \nonumber\\
&+& \left(N(\bar U_{k|l}+\bar U_{l|k})+(\bar N_{k|l}+\bar N_{l|k})\hat{D}\right)\frac{\partial\tilde U}{\partial\gamma_{kl}}
\nonumber\\
&+& (N u)_{,i}\bar U^i\frac{\partial\tilde U}{\partial u}-\left(\bar N^k\bar U^i_{|k}-\bar U^k \bar N^i_{|k}+\gamma^{ik}(N u)_{,k} \right)\frac{\partial\tilde U}{\partial u^i}\nonumber\\
&-& \left(
\bar N^i\hat{D}\tilde U
+N u\gamma^{ik}\frac{\partial\tilde U}{\partial u^k}
+2N\bar U^j\gamma_{jk}\frac{\partial\tilde U}{\partial\gamma_{ik}} 
\right)_{,i}=0.\label{eq:Omt}
\end{eqnarray}
Therefore, we have got a new (tertiary) constraint. Given zero Dirac bracket $\{\tilde\Omega,\pi_u\}_{\tilde D}$, this constraint does not depend on variable $u$ when ${\cal S}_i$ constraints are taken into account.  Eq.~(\ref{eq:Omt}) is linear in momenta $\pi^{ij}$ and so can be solved for the momentum variable which is conjugate to the ghost variable.

Next step is to derive the Hamiltonian equation for constraint $\tilde\Omega$ 
$$
\dot{\tilde\Omega}=\{\tilde\Omega, {\rm H_{reduced}} \}_{\tilde D}=\tilde\Psi,
$$
and to require that $\dot{\tilde\Omega}$ should be zero. We note that $\tilde\Psi$ cannot depend on Lagrangian multiplier $\lambda$, because $\{\tilde\Omega,\pi_u\}_{\tilde D}=0$, similar to Eq.(\ref{eq:ThetaOmega}). So we get the last constraint $\tilde\Psi$, which serves for fixing variable $u$. At last, $\lambda$ is fixed by requirement $\dot{\tilde\Psi}=\{\tilde\Psi,{\rm H_{reduced}}\}=0$, compare (\ref{eq:OS}), (\ref{eq:VHH}).
At the end, we have 4 second class constraints ${\cal S}$, $\tilde\Omega$, $\pi_u$, $\tilde\Psi$ for 7 pairs of gravitational canonical variables. Two second class constraints $\tilde \Psi$ and $\pi_u$ serve to kill a pair of canonical variables $(u,\pi_u)$. Two other second class constraints serve to kill the ghost degree of freedom. Therefore this gives $5$ degrees of freedom for massive gravity.  
The final Hamiltonian can be written as follows
$$
{\rm H_{final}}=\int d^3x\left(N\left(\tilde U -u\frac{\partial\tilde U}{\partial u} - u^i \frac{\partial\tilde U }{\partial u^i}\right)-N^i\frac{\partial\tilde U}{\partial u^i}\right),
$$
and this is the same formula as Eq.(70) derived in~\cite{SolTch} for the potential of a general form.

\section{Conclusion}
We have studied here the canonical structure of a model  which, as we hope, should be isomorphic to the widely discussed Hassan-Rosen bigravity~\cite{HR_bi} or, if one spacetime metric is a fixed solution of GR equations, to the dRGT theory~\cite{dRGT} of massive gravity. We found that in order to obtain a ghost-free bigravity theory satisfying  diagonal 4-diffeomorphism invariance it is sufficient to take a  potential  fulfilling the following 4 conditions: 
\begin{enumerate}
\item  the potential can be expressed as a function of variables which are components of $3+1$-decomposed two metrics $\tilde U(u,u^i,\eta_{ij},\gamma_{ij})$; 
\item  the potential allows us to have 4 first class constraints ${\cal R},{\cal R}_i$ in the bigravity case; 
\item  the Hessian of the potential in the lapse-like and shift-like variables $u,u^i$ is degenerate\footnote{Our variables can be expressed through lapses and shifts of the two metrics  as follows $u=\bar N/N$, $u^i=(\bar N^i- N^i)/N$; see Eq.(\ref{eq:uui}).};
\item  the Hessian in the shift-like  variables $u^i$ is nondegenerate. 
\end{enumerate}
It would be an interesting problem to clear whether the dRGT potential~\cite{dRGT} is a unique realization of the above axioms.

Since the first part of this work~\cite{SolTch} has appeared, some interesting and related articles have been published. Kluson's preprint~\cite{Kluson:2012ps}, in fact, was simultaneous to~\cite{SolTch}, and the same algebra of first class constraints in bigravity was proved under some conditions imposed on the potential, the dRGT case was not included. Distinctions from our work are in a narrower class of potentials, in a choice of variables, and in using ADM approach, not Kucha\u{r}'s one.    
Alexandrov et al~\cite{Krasnov} developed Ashtekar-like approach with the corresponding variables. These authors were careful to mention some difficulties with the reality conditions. Their main conclusion is a confirmation of statements made in  articles~\cite{HR,HR093230,HR_bi,HR2},  at least for one variant of the dRGT potential. Many interesting ideas are suggested, including a choice of an arithmetic average of triads related to different metrics.  

Two other articles by Kluson~\cite{Kluson:2013,Kluson:2013-2} have appeared later and contain some pessimistic conclusions related to  bigravity. Contrary to our results, it is claimed that a constraint playing the role analogous to ${\cal R}$ in the present work becomes second class and so the theory loses the diagonal diffeomorphism invariance. As we have shown here, ${\cal R}$ stays the first class constraint, as does its analog in GR. 
We think that some further calculations may remove the mentioned difficulty. See, for example footnote 5 of Ref.~\cite{Kluson:2013-2}, showing that, in fact, there is a space for the explicit check of some relations considered in Ref.~\cite{Kluson:2013-2} as ``highly improbable''. 

At last, work by Comelli {\it et al.}~\cite{Comelli2013, Comelli_MayDay} seems most important for us. This group has independently started a study of  massive gravity (they do not consider the bigravity case, which is a main object of our work) with the potential of a general form and we have been able to compare our preliminary results with those announced in Ref.~\cite{Comelli2013}. A detailed  presentation~\cite{Comelli_MayDay} has appeared after submission of this work. 
If one wants to compare the conditions imposed on the potential in Refs.~\cite{Comelli2013, Comelli_MayDay} and in the present work, it is easy to see the difference. Our second axiom is absent in their scheme. The price, of course, should be paid: it is necessary for them to treat Eq.~(\ref{eq:Theta}) proved in our work as an independent axiom. 
It is important to note that this group has also proposed a new class of  massive gravity theories which is not only free of ghosts but has such phenomenologically attractive features as weak coupling and high ultraviolet cutoff.

{\bf Acknowledgements} We are pleased to thank S.~Deser, S.F.~Hassan, M.V.~Neshchadim, L.~Pilo, A.V.~Razumov, S.~Speziale, A.~Waldron, and  Yu.M.~Zinoviev for comments and discussions. One of the authors (V.O.S.) is grateful to the organizers and participants of Workshop on Infrared Modifications of Gravity  (ICTP, Trieste, September 26 -- 30 , 2011) for stimulating atmosphere and to Prof. S. Randjbar-Daemi for hospitality during his visit to ICTP.

\appendix

\section{Calculating Poisson brackets of ${\cal R}$, ${\cal R}_i$}\label{S:FC}
In calculating Poisson brackets between constraints ${\cal R}$, ${\cal R}_i$ we can treat 
$u,u^i$
as functions, and not as canonical variables, because their conjugate momenta do not appear in these relations. Then potential $\tilde U$ has nonzero Poisson brackets only with gravitational momenta $\Pi^{ij},\pi^{ij}$, this results in appearance of derivatives $\partial\tilde U/\partial\eta_{ij}$ and $\partial\tilde U/\partial\gamma_{ij}$.
In derivation of the first bracket
\begin{equation}
  \{ {\cal R}_i(x),{\cal R}_j(y)\}={\cal R}_i(y)\delta_{,j}(x,y)+ {\cal R}_j(x)\delta_{,i}(x,y),\label{eq:bialg2}
\end{equation}
we take into account that ${\cal H}_i$ and $\bar{\cal H}_i$ commutes, and each of them satisfies 
Eq.~(\ref{eq:alg2}), so (\ref{eq:bialg2}) is valid.
By straightforward calculations we get
\begin{eqnarray}
\{{\cal R}(x),{\cal R}(y)\}&=&\left(
\eta^{ij}{\cal R}_j+uu^i{\cal S}-(\eta^{ij}-u^2\gamma^{ij}-u^i u^j ){\cal S}_j+Q^i
\right)(x)\delta_{,i}(x,y)\nonumber\\
&-&\left(
\eta^{ij}{\cal R}_j+uu^i{\cal S}-\left(\eta^{ij}-u^2\gamma^{ij}-u^i u^j \right){\cal S}_j+Q^i
\right)(y)\delta_{,i}(y,x)\nonumber\\
&\approx& \left(
\eta^{ij}{\cal R}_j+Q^i
\right)(x)\delta_{,i}(x,y)-\left(
\eta^{ij}{\cal R}_j+Q^i
\right)(y)\delta_{,i}(y,x),\label{eq:RR}
\end{eqnarray}
and also,
\begin{eqnarray}
 \{{\cal R}_i(x),{\cal R}(y)\}&=& {\cal R}(x)\delta_{,i}(x,y)+u_{,i}{\cal S}\delta(x,y)\nonumber\\
&+&\frac{\partial}{\partial x^j}\left(Q^j_i(x)\delta(x,y) \right)\nonumber\\
&+&\frac{\partial}{\partial x^j}\left(u^j {\cal S}_i(x)\delta(x,y)\right)+u^j_{,i}{\cal S}_j\delta(x.y)\nonumber\\
&\approx&  {\cal R}(x)\delta_{,i}(x,y)+\frac{\partial}{\partial x^j}\left(Q^j_i(x)\delta(x,y) \right),\label{eq:RRi}
\end{eqnarray}
where simpler expressions arise after taking into account constraint equations  ${\cal S}\approx 0$ and ${\cal S}_i\approx 0$. Quantities
\begin{equation}
{ Q}^i_k= 2\eta_{jk}\frac{\partial\tilde{U}}{\partial\eta_{ij}}+2\gamma_{jk}\frac{\partial\tilde{U}}{\partial\gamma_{ij}}-u^i\frac{\partial\tilde{U}}{\partial u^k}-\delta^i_k\tilde{U},
\end{equation}
\begin{equation}
{ Q}^\ell= 2u^j\gamma_{jk}\frac{\partial\tilde{U}}{\partial\gamma_{k\ell}}-u^\ell u\frac{\partial\tilde{U}}{\partial u}+\left(\eta^{k\ell}-u^2\gamma^{k\ell}-u^k u^\ell \right)\frac{\partial\tilde{U}}{\partial u^k},
\end{equation}
should be equal to zero in order  ${\cal R}$, ${\cal R}_i$ to be first class constraints.

As has been observed in articles~\cite{Krasnov,Kluson:2012ps}, Eq.~(\ref{eq:RRi}) becomes simpler if one add to ${\cal R}_i$ a combination of primary constraints:
\begin{equation}
 \tilde{\cal R}_i={\cal R}_i+\pi_u u_{,i}+\pi_{u^k}u^k_{\ ,i}+(\pi_{u^i}u^k)_{,k},
\end{equation}
then we obtain
\begin{equation}
 \{\tilde{\cal R}_i(x),{\cal R}(y)\} ={\cal R}(x)\delta_{,i}(x,y)+\frac{\partial}{\partial x^j}\left(Q^j_i(x)\delta(x,y) \right),
\end{equation}
but with the replacement of ${\cal R}_i$ for $\tilde{\cal R}_i$ the primary constraints appear in 
Eq.~(\ref{eq:RR}).

\section{Solving ${\cal S}_i$ constraints}\label{solv_const}
Second class constraints may be used not only for deriving the Dirac brackets, but also in the general case for straightforward exclusion of conjugate variables  $(u^i,\pi_{u^i})$ from the Hamiltonian.  When the chosen potential is simple enough (see, for example,~\cite{SolTch2011}), it is possible to express 3 functions $u^i$ (or $g^{\perp i}$) through the rest variables after solving  constraints 
\begin{equation}
\bar{\cal H}_i+\partial\tilde U/\partial u^i=0.
\end{equation} 
Given these solutions
\begin{equation}
 u^k=\bar u^k(u,\eta_{ij},\gamma_{ij},\bar{\cal H}_i),\label{eq:solve_ui}
\end{equation}
and then substituting them into constraint $\bar{\cal H}+\partial\tilde U/\partial u=0$, we can express also $u$ (or $g^{\perp\perp}$) as a function of the rest variables
\begin{equation}
u=\bar u(\eta_{ij},\gamma_{ij},\bar{\cal H}(\gamma_{ij},\pi^{ij},\phi_A,\pi^A)).\label{eq:usolved}
\end{equation}

It is stated that for potential of a special form proposed and studied in articles~\cite{dRGT,HR} it is impossible to solve the secondary constraints (\ref{eq:last}) and express $u$ as a function (\ref{eq:usolved}).
As above, we suppose that potential $\tilde U$ has such a property that 3 constraints ${\cal S}_i$ can be solved for variables $u^k$, matrix $||L_{ij}||$ is invertible in this case. By substituting these solutions back into equations ${\cal S}_i=0$ we evidently get identities, and then
\begin{equation}
\frac{\partial}{\partial u}\left.{\cal S}_i\right|_{u^k=\bar u^k}\equiv 0,\label{eq:ident}
\end{equation}
or equivalently,
\begin{equation}
\frac{\partial^2\tilde U}{\partial u\partial u^i}+\frac{\partial^2\tilde U}{\partial u^i\partial u^k}\frac{\partial \bar u^k}{\partial u}= 0.
\end{equation}
From Eq.(\ref{eq:ident}) and from analogous equations with other derivatives of ${\cal S}$ we find
\begin{eqnarray}
\frac{\partial\bar u^k}{\partial u}&=&-(L^{-1})^{k\ell}\frac{\partial^2\tilde U}{\partial u\partial u^\ell}=-\bar{U}^k\qquad \frac{\partial\bar u^k}{\partial\bar{\cal H}_i}=-\left(L^{-1}\right)^{ik},\\
\frac{\partial\bar u^k}{\partial\eta_{mn}}&=&-\left(L^{-1}\right)^{ik}\frac{\partial^2\tilde U}{\partial u^i\partial\eta_{mn}},\qquad \frac{\partial\bar u^k}{\partial\gamma_{mn}}=-\left(L^{-1}\right)^{ik}\frac{\partial^2\tilde U}{\partial u^i\partial\gamma_{mn}}.
\end{eqnarray}
Next, after substitution of the derived expressions 
$u^k=\bar u^k(u,\eta_{ij},\gamma_{ij},\bar{\cal H}_i)$ into the forth constraint, we get an expression, which according to supposition made in articles~\cite{dRGT,HR} does not depend on $u$:
\begin{equation}
{\cal S}_{\rm reduced}=\bar{\cal H}+\left.{\frac{\partial \tilde U}{\partial u}}\right|_{u^k=\bar u^k},
\end{equation}
and then
\begin{equation}
\frac{\partial {\cal S}_{\rm reduced}}{\partial u}=\frac{\partial^2\tilde U}{\partial u^2}+\frac{\partial \bar u^i}{\partial u}\frac{\partial^2\tilde U}{\partial u\partial u^i}=\hat{D}\frac{\partial\tilde U}{\partial u}=0.
\end{equation}

In order to commute the constraints 
we can use equivalently the canonical Poisson brackets instead of Dirac brackets, if we take into account dependence given by Eq.~(\ref{eq:solve_ui}).

\section{Analysis of bigravity with Lagrangian multipliers method}\label{Lagr_mult}
Apart from the two methods used above (Dirac brackets construction in Section 3 and solving of constraints in Appendix~\ref{solv_const}) there is an approach based on calculation of the Lagrangian multipliers. Here we start with the Hamiltonian containing all primary constraints with their Lagrangian multipliers, i.e. functions which should be varied to get constraints as equations of motion, but which have zero Poisson brackets with any variable,
\begin{equation}
{\rm H}=\int 
\left(
N{\cal R}+N^i{\cal R}_i+\lambda\pi_u+\lambda^i\pi_{u^i}  
\right).
\end{equation}
Then we check for consistency 4 primary constraints:
\begin{eqnarray}
\dot\pi_u&=&\{
\pi_u,{\rm H}
\}=\int d^3x 
N\{\pi_u,{\cal R}\}=-\frac{\partial{\cal R}}
{\partial u}=-N{\cal S}=0,\\
\dot\pi_{u^i}&=&\{\pi_{u^i},{\rm H}\}=\int d^3xN\{\pi_{u^i},{\cal R}\}=-\frac{\partial{\cal R}}{\partial u^i}=-N{\cal S}_i=0,
\end{eqnarray}
and get 4 secondary constraints
\begin{equation}
{\cal S}\approx 0,\qquad {\cal S}_i\approx 0.
\end{equation}
We should not add secondary constraints to the Hamiltonian, but we are  to check their evolution equations for the consistency
\begin{eqnarray}
\dot{\cal S}&=&
\int 
N\{{\cal S},{\cal R}\}+N^k\{{\cal S},{\cal R}_k\}+\lambda\{{\cal S},\pi_u\}+\lambda^k\{{\cal S},\pi_{u^k}\}  
=0,
\label{eq:nolambda}\\
\dot{\cal S}_i&=&
\int 
N\{{\cal S}_i,{\cal R}\}+N^k\{{\cal S}_i,{\cal R}_k\}+\lambda\{{\cal S}_i,\pi_u\}+\lambda^k\{{\cal S}_i,\pi_{u^k}\}  
=0,\label{eq:Lik}
\end{eqnarray}
where omitting $\delta$-function we can write
\begin{equation}
\{{\cal S}_i,\pi_{u^k}\}=L_{ik},\quad \{{\cal S}_i,\pi_u\}=\tilde U_i=\{{\cal S},\pi_{u^i}\},\quad \{{\cal S},\pi_u\}=\tilde U''.  
\end{equation}
As $L_{ik}$ is invertible  matrix we can solve Eq.~(\ref{eq:Lik}) for  $\lambda^k$
\begin{equation}
\lambda^k=-\lambda \bar U^k-(L^{-1})^{kj}\left(N\{{\cal S}_j,{\cal R}\}+N^i\{{\cal S}_j,{\cal R}_i\}\right),
\end{equation}
and then substitute this into Eq.~(\ref{eq:nolambda}). Evidently, multiplier $\lambda$ is canceled due to Eq.~(\ref{eq:L}), and therefore Eq.~(\ref{eq:nolambda}) is a new constraint (compare with 
Eq.~(\ref{eq:Omega})):
\begin{equation}
\dot{\cal S}\approx N\Omega\approx 0,\qquad \mbox{or} \qquad \Omega\approx 0.
\end{equation}
Next step is to check a consistency of this last constraint
\begin{eqnarray}
\dot\Omega&=&\int 
N\{\Omega,{\cal R}\}+N^k\{\Omega,{\cal R}_k\}+\lambda\{\Omega,\pi_u\} +\lambda^k\{\Omega,\pi_{u^k}\}\nonumber\\
&=& \int N\left[ \{\Omega,{\cal R}\}-\{\Omega,\pi_{u^i}\}(L^{-1})^{ij}\{{\cal S}_j,{\cal R}\}\right]\nonumber\\
&+& N^i\left[ \{\Omega,{\cal R}_i\}-\{\Omega,\pi_{u^k}\}(L^{-1})^{kj}\{{\cal S}_j,{\cal R}_i\}\right]\nonumber\\
&+&\lambda\left[\{\Omega,\pi_u\}-\bar U^i\{\Omega,\pi_{u^i}\}\right]\nonumber\\
&=&\int N \{\Omega,{\cal R}\}_{\tilde D}+N^i \{\Omega,{\cal R}_i\}_{\tilde D}+\lambda \{\Omega,\pi_u\}_{\tilde D}\approx 0.\label{eq:dotOmega}
\end{eqnarray}
We can see from Eq.~(\ref{eq:ThetaOmega}) that the last Dirac bracket is proportional to $\Theta^i$,  so
 it is zero.   Then Eq. (\ref{eq:dotOmega}) gives a new constraint. As $\Omega$ is a spatial scalar we have 
\begin{equation}
 \{\Omega,{\cal R}_k\}_{\tilde D}\approx 0,
\end{equation}
and therefore the new condition is as follows
\begin{equation}
\int N\{\Omega,{\cal R}\}_{\tilde D}=\Psi\approx 0.
\end{equation}
Next we should check
\begin{eqnarray}
 \dot\Psi&=&\{\Psi,{\rm H}\}=\int 
N\{\Psi,{\cal R}\}+N^k\{\Psi,{\cal R}_k\}+\lambda\{\Psi,\pi_u\}+\lambda^k\{\Psi,\pi_{u^k}\}\nonumber\\
&=&\int 
N\{\Psi,{\cal R}\}_{\tilde D}+N^k\{\Psi,{\cal R}_k\}_{\tilde D}+\lambda\{\Psi,\pi_u\}_{\tilde D}=0.
 \label{eq:lastlast}
\end{eqnarray}
As, according to Eqs.~(\ref{eq:OS}), (\ref{eq:VHH}),
\begin{equation}
 \{\Psi,\pi_u\}_{\tilde D} \ne 0,
\end{equation}
 Eq. (\ref{eq:lastlast}) can be solved for $\lambda$ and the analysis of constraints is finished.

\section{Dictionary}\label{S:MayDay}
Besides the present work the program to study the massive gravity and the bigravity starting from the general potential is carried out independently  by at least two other groups. Comelli, Nesti, Pilo~\cite{Comelli2013,Comelli2012,Comelli_MayDay} pay their attention to massive gravity case and Kluson~\cite{Kluson:2012ps,Kluson:2013,Kluson:2013-2} concentrates on the bigravity. As each group uses a lot of their own notations we think it would be useful to have dictionaries to translate between them. We are adding this Appendix in order to make easier the reading of one paper for someone who is already familiar with another. 
\pagestyle{empty}
\renewcommand{\hoffset}{\hoffset=-50mm}
\begin{table}
\setlength{\extrarowheight}{5pt}
\begin{tabular}{|c||c|c|}
\hline 
variables and equations & this work & Comelli, Nesti and Pilo \\
\hline
\hline
GR 4-metric & $g_{\mu\nu}$ & $g_{\mu\nu}$ \\
\hline
additional 4-metric  & $f_{\mu\nu}$ & ${\tilde g}_{\mu\nu} $\\
\hline
GR 3-metric & $\gamma_{ij}$ & $\gamma_{ij}$ \\
\hline
conjugate momenta & $\pi^{ij}$ & $\Pi^{ij}$\\
\hline 
GR constraints & $(\bar{\cal H},\bar{\cal H}_i)$ & $({\cal H},{\cal H}_i)={\cal H}_A$\\
\hline
additional 3-metric & $\eta_{ij}$ & $f_{ij}$ or $\delta_{ij}$ \\
\hline
additional momenta & $\Pi^{ij}$ & no \\
\hline
GR lapse and shift & $(\bar N,{\bar N}^i)$ & $(N, N^i)=N^A$ \\
\hline
additional lapse-shift & $(N,{N}^i)$ & $(1,0)$ \\
\hline
combined variables & $u={\bar N}/N$, $u^i=({\bar N}^i-N^i)/N$ &  no \\
\hline
primary constraints & $(\pi_u,\pi_{u^i}),(\pi_N,\pi_{N^i})$ & $\Pi_A$\\
\hline
secondary constraints & $({\cal S},{\cal S}_i),({\cal R},{\cal R}_i)$ & $({\cal S}_0,{\cal S}_i)={\cal S}_A$ \\
\hline
tertiary constraint &$\Omega$& $({\cal T}_0,{\cal T}_i)={\cal T}$\\
\hline
quaternary constraint &$\Psi$& ${\cal Q}$\\
\hline
potential in action & $\sqrt{-f}U=N\sqrt{\eta}U$ & $m^2{M_{pl}}^2\sqrt{|g|}V$\\
\hline
potential in hamiltonian & $N\tilde U(u^a)$ &${\cal V}(N^A)=m^2N\sqrt{\gamma}V$ \\
\hline
potential as scalar & $NU$ & $\tilde{\cal V}= \gamma^{-1/2}{\cal V}$ \\
\hline
second derivatives matrix & $
\left(
\begin{array}{cc}
 \tilde U''
& \tilde U_j \\
\tilde U_i
& 
L_{ij}(x) 
\end{array}
\right)=\frac{\partial^2{\tilde U}}{\partial u^a\partial u^b}$ & ${\cal V}_{AB}=\frac{\partial^2{\cal V}}{\partial N^A\partial N^B}$ \\
\hline
Monge-Amp\`ere equation & $\left|\frac{\partial^2{\tilde U}}{\partial u^a\partial u^b}\right|=0$& $\left|\frac{\partial^2{\cal V}}{\partial N^A\partial N^B}\right|=0$ \\
\hline
null vector & $(1,-(L^{-1})^{ij}\tilde U_j)=(1, -\bar U^i)$ & $(1, -{\cal V}^{-1}_{ij}{\cal V}_{0j})=\chi^A$\\
\hline
null derivative & $\hat D=\frac{\partial}{\partial u}-\bar U^i\frac{\partial}{\partial u^i}$ & $\chi^A\partial_A=\chi^A\frac{\partial}{\partial N^A}$\\
\hline
key result no.1 & $\{{\cal S},{\cal S}\}+\bar U^i\{{\cal S}_i,{\cal S}_j\}\bar U^j\approx \Theta^i\delta_{,i}$ & $\chi^A\{{\cal S}_A,{\cal S}_B\}\chi^B\approx \Theta$ \\
\hline
key equation $\Theta^i=0$ & $\left(\bar U^k\hat{D}\left(\delta^i_k-2\gamma_{jk}\frac{\partial}{\partial\gamma_{ij}} \right)-\gamma^{ij}\frac{\partial}{\partial u^j}\right)\tilde U=0$ & ${\chi^0}^2\tilde{\cal V}_i  + 2\chi^A  \chi^j\frac{\partial \tilde{\cal V}_A}{\partial \gamma^{ij}}=0$\\
\hline
key result no.2 & $\{\Psi,\pi_u\}-\bar U^i\{\Psi,\pi_{u^i}\}=\hat D\Psi\ne 0$ & $\chi^A{\cal Q}_{,A}\ne 0$ \\
\hline
final Lagrangian multiplier & $\lambda$ & $z$ \\
\hline
final massive gravity H & $\int d^3x\left( 
N(\tilde U-u\frac{\partial\tilde U}{\partial U}-u^i\frac{\partial\tilde U }{\partial u^i})-N^i\frac{\partial\tilde U }{\partial u^i}\right)$ & $\int d^3x \left({\cal V}-{\cal V}_A N^A\right)$\\
\hline
final bigravity H & $\int d^3x\left(N{\cal R}+N^i{\cal R}_i\right)$ & no\\
\hline
\end{tabular}
\caption{Dictionary to translate between this work notations and notations of Ref.~\cite{Comelli_MayDay} (part 1)}
\end{table}

\begin{table}
\setlength{\extrarowheight}{5pt}
\begin{tabular}{|c||c|c|}
\hline 
variables and equations & this work & Comelli, Nesti and Pilo \\
\hline
\hline
arbitrary null vector & $(\chi^0,-\chi^0\bar U^i)$ & $\chi^A=(\chi^0,\chi^i)$\\
\hline
new variable & $\xi^i=-\bar U^i$ & $\xi^i\equiv\frac{\chi^i}{\chi^0}$ \\
\hline
new function & $\chi^0\hat DU$ & $\mathbf{U}(\chi^A)\equiv\tilde{\cal V}_A\chi^A$ \\
\hline
another new function & $ V(\xi^i)\equiv\hat D\tilde U$ & ${\cal V}_A\chi^A/\chi^0$ \\
\hline
homogeneity & trivial & ${\bf U}(\kappa\chi^A)=\kappa{\bf U}(\chi^A)$ \\
\hline
new function  & trivial & $\mathbf{U}(\chi^A)=\chi^0 {\cal U}(\xi^i)$ \\
\hline
this new function &${ V}(\xi^i)(\sqrt{\eta})^{-1}$& ${\cal U}(\xi^i)$\\
\hline
scalar derivative & $\frac{\partial U}{\partial u}={ V}(\sqrt{\eta})^{-1}-\frac{\partial{ V(\sqrt{\eta})^{-1}}}{\partial\xi^j}\xi^j$ & $\tilde{\cal V}_0={\cal U}-\frac{\partial{\cal U}}{\partial\xi^j}\xi^j$ \\
\hline 
density derivative& $\frac{\partial \tilde U}{\partial u}={ V}-\frac{\partial{ V}}{\partial\xi^j}\xi^j$ & no \\
\hline 
scalar derivatives & $\frac{\partial U}{\partial u^i}=\frac{\partial V(\sqrt{\eta})^{-1}}{\partial\xi^i}$ & $\tilde{\cal V}_i=\frac{\partial{\cal U}}{\partial\xi^i}$ \\
\hline
density derivatives & $\frac{\partial \tilde U}{\partial u^i}=\frac{\partial V}{\partial\xi^i}$ & no \\
\hline
integrability condition & $\hat D\xi^j\equiv \frac{\partial \xi^j}{\partial u}+\xi^k\frac{\partial\xi^j}{\partial u^k}=0$ & $\frac{\partial\xi^j}{\partial N}+\frac{\partial\xi^j}{\partial N^k}\xi^k=0$\\
\hline 
its solution & $u^i=u\xi^i + (V^{-1})^{ik}\frac{\partial W}{\partial\xi^k}$ & $N^i=N\xi^i+{\cal Q}^i$ \\
\hline
new expression & $(V^{-1})^{ik}\frac{\partial W}{\partial\xi^k}$ & ${\cal Q}^i=-({\cal U}_{ij})^{-1}\frac{\partial{\cal E}}{\partial\xi^j}$ \\
\hline 
new density & $-W=\tilde U-u\frac{\partial\tilde U}{\partial u}-u^i\frac{\partial\tilde U}{\partial u^i}$ & $\sqrt{\gamma} {\cal E}$ \\
\hline
corresponding scalar & $-(\sqrt{\eta})^{-1}W$ & ${\cal E}=\tilde {\cal V}-N{\cal U}-{\cal U}_i{\cal Q}^i$\\
\hline
potential as scalar & $U$ & $\tilde {\cal V}=N{\cal U}+{\cal U}_i{\cal Q}^i+{\cal E}$\\
\hline
potential as density & $\tilde U=uV-W+(V^{-1})^{ij}\frac{\partial V}{\partial\xi^i}\frac{\partial W}{\partial\xi^j}$ & ${\cal V}$ \\
\hline
energy density & $- W$ & $\sqrt{\gamma} {\cal E}$ \\
\hline
\end{tabular}
\caption{Dictionary to translate between this work notations and notations of Ref.~\cite{Comelli_MayDay} (part 2)}
\end{table}

\begin{table}
\setlength{\extrarowheight}{5pt}
\begin{tabular}{|l||l|l|}
\hline 
notations & this work & Kluson \\
\hline
\hline
GR 4-metric & $g_{\mu\nu}$ & $\hat g_{\mu\nu}$ \\
\hline
additional 4-metric  & $f_{\mu\nu}$ & ${\hat f}_{\mu\nu} $\\
\hline
GR 3-metric & $\gamma_{ij}$ & $g_{ij}$ \\
\hline
conjugate momenta & $\pi^{ij}$ & $\pi^{ij}$\\
\hline 
additional 3-metric & $\eta_{ij}$ & $f_{ij}$  \\
\hline
additional momenta & $\Pi^{ij}$ & $\rho^{ij}$ \\
\hline
GR lapse and shift & $\bar N,{\bar N}^i$ & $N, N^i$ \\
\hline
additional lapse-shift & $N,{N}^i$ & $M,L^i$ \\
\hline
combined variables & $u={\bar N}/N$, $u^i=({\bar N}^i-N^i)/N$ &  $n=\sqrt{N/M}$, $n^i=(N^i-L^i)/\sqrt{NM}$ \\
\hline
combined variables (2) &  &  $\bar N=\sqrt{NM}$, $\bar N^i=(N^i+L^i)/2$ \\
\hline
conjugate momenta &$(N,\pi_N)$, $(N^i,\pi_{N^i})$,  & $(\bar N,P_{\bar N})$, $(\bar N^i,P_i)$,  \\
&$(u,\pi_u)$, $(u^i,\pi_{u^i})$ & $(n,p_n)$, $(n^i,p_i)$\\
\hline
combinations & $\bar N=uN$ & $N=\bar N n$, $M=\bar N/n$\\
\hline
combinations (2) & $\bar N^i=N^i+u^iN$ & $L^i=\bar N^i-n^i\bar N/2$, $N^i=\bar N^i+n^i\bar N/2$ \\
\hline
change of variables & $\bar N^i-N^i=\bar N\xi^i + N(V^{-1})^{ik}\frac{\partial W}{\partial\xi^k}$ & $N^i-L^i=M\tilde n^i+N\tilde D^i_{\ j}{\tilde n}^j$ \\
\hline
primary constraints & $(\pi_u,\pi_{u^i}),(\pi_N,\pi_{N^i})$ & $(P_N,P_i),(p_n,p_i)$\\
\hline
secondary constraints & $({\cal S},{\cal S}_i),({\cal R},{\cal R}_i)$ & $({\cal G}_n,{\cal G}_i),(\bar{\cal R},\bar{\cal R}_i)$ \\
\hline
GR constraints & $(\bar{\cal H},\bar{\cal H}_i)$ & (${\cal R}_0^{(g)}$,${\cal R}_i^{(g)}$) \\
\hline
potential in action & $\sqrt{-f}U$ & $\mu\sqrt[4]{\hat g\hat f}{\cal V}$ \\
\hline
time translation & ${\cal H}+u\bar{\cal H}+u^i\bar{\cal H}_i+\tilde U$ & ${\cal R}_0^{(f)}/n+n{\cal R}_0^{(g)}$\\
& & $+n^i({\cal R}_i^{(g)}-{\cal R}_i^{(f)})/2+\mu^2\sqrt[4]{fg}{\cal V}$\\
\hline
3-diff generator & ${\cal R}_i={\cal H}_i+\bar{\cal H}_i$ & $\bar{\cal R}_i={\cal R}^{(f)}_i+{\cal R}_i^{(g)}$\\
\hline
revised & & $\tilde{\cal R}_i=\bar{\cal R}_i+p_n\partial_i n+p_j\partial_i n^j+\partial_j(n^jp_i)$\\
\hline
Hamiltonian density & $N{\cal R}+N^i{\cal R}_i$ & $\bar N\bar{\cal R}+\bar N^i\bar{\cal R}_i$\\
\hline
additional terms & $\lambda\pi_u+\lambda^i\pi_{u^i}+\lambda^N\pi_N+\lambda^{N^i}\pi_{N^i}$ & $V_NP_{\bar N}+V^iP_i+{v}_np_n+{v}^ip_i+u^n{\cal G}_n+u^i{\cal G}_i$ \\
\hline
second class 
constraints 
& $\pi_a=(\pi_u,\pi_{u^i}),{\cal S}_b=({\cal S},{\cal S}_i)$ & $p_a=(p_n,p_i),{\cal G}_b=({\cal G}_n,{\cal G}_i)$\\
\hline
their Poisson 
brackets 
& $
\left(
\begin{array}{cc}
0
& -\tilde U_{ab} \\
\tilde U_{ab}
& 
 \{ {\cal S}_{a},{\cal S}_b\} 
\end{array}
\right)$ &  $
\left(
\begin{array}{cc}
0
& \Delta_{ab} \\
-\Delta_{ab}
& 
\{ {\cal G}_{a},{\cal G}_b\} 
\end{array}
\right)$\\
\hline
degeneration & ${\rm Det}\tilde U_{ab}\equiv \left|\frac{\partial^2{\tilde U}}{\partial u^a\partial u^b}\right|=0$ & ${\rm Det}\Delta_{ab}=0$ \\
\hline
null vector & $(1,-(L^{-1})^{ij}\tilde U_j)=(1, -\bar U^i)$ & $\Delta_{ab}u^b_0=0$\\
\hline
new notations & & $\tilde p=u^a_0p_a, \tilde{\cal G}=u^a_0{\cal G}, u^a_0\tilde p_a=0, u^a_0\tilde {\cal G}_a=0$\\
\hline
separation of 
constraints 
& $(\pi_u,{\cal S}),(\pi_{u^i},{\cal S}_j)$ & $(\tilde p,\tilde{\cal G}),(\tilde p_a,\tilde {\cal G}_b)$\\
\hline
\end{tabular}
\caption{Dictionary to translate between this work notations and notations of Ref.~\cite{Kluson:2013-2}}
\end{table}

\begin{thebibliography}{**}



\bibitem{Rosen}
 N.~Rosen, 
%General Relativity and flat space. I,II.
{\it Phys. Rev.} %Vol.
{\bf 57} 147-150; 150-153 (1940);
%Flat-space metric in general relativity theory.
{\it Ann. of Phys.} {\bf 22}, 1-11 (1963);
%A bi-metric theory of gravitation.
{\it Gen. Rel. Grav.} {\bf 4}, 435-447 (1973).

\bibitem{Salam} 
%A.~Salam and J.~Strathdee, {\it Phys. Rev.} {\bf 184} 1750-1759; 1760-1768 (1969); 
C.~J.~Isham, A.~Salam and J.~Strathdee, {\it Phys. Lett. B} {\bf 31} 300-302 (1970); {\it Phys. Rev.} {\bf D3} 867-873 (1971).

\bibitem{Kogan}
T. Damour and I.I. Kogan,
{\it Phys.Rev. } {\bf D 66}  104024 (2002).

\bibitem{BD}
D.G. Boulware and S. Deser, {\it Phys.Rev. } {\bf D6} 3368-3382 (1972). 
%(Can gravitation have a finite range?)

\bibitem{Deser}
S.~Deser, A.~Waldron. %Acausality of massive gravity. 
{\it Phys. Rev. Lett.} {\bf 110}, 111101 (2013); arXiv:1212.5835;
S. Deser, K. Izumi, Y. C. Ong, A. Waldron. Superluminal Propagation and Acausality of Nonlinear Massive Gravity; arxiv:1312.1115.

\bibitem{dRGT}
C. de Rham, G. Gabadadze, A. J. Tolley,
% Resummation of Massive Gravity,
{\it Phys. Rev. Lett.} {\bf 106} 231101 (2011); 	arXiv:1011.1232;
% Ghost free Massive Gravity in the Stückelberg language,
{\it Phys. Lett. B}
%Volume 
{\bf 711} 190–195 (2012); arXiv:1107.3820

\bibitem{WZ} B.~Zumino, ``Effective Lagrangians and broken symmetries,''
  in Brandeis Univ. Lectures on Elementary Particles and Quantum Field Theory (MIT Press Cambridge, Mass.), Vol. 2,  1970, 437.

\bibitem{DM}
G. D'Amico, C. de Rham, S. Dubovsky, G. Gabadadze, D. Pirtskhalava, A.J. Tolley.
{\it Phys. Rev. D} {\bf 84}, 124046 (2011);
%Massive Cosmologies,
	arXiv:1108.5231.

\bibitem{HR}
S. F. Hassan, Rachel A. Rosen, 
%Resolving the Ghost Problem in non-Linear Massive Gravity,
{\it Phys. Rev. Lett.} {\bf 108} 041101 (2012),	arXiv:1106.3344.

\bibitem{HR093230}
	S. F. Hassan, Rachel A. Rosen, Angnis Schmidt-May,
	%Ghost-free Massive Gravity with a General Reference Metric,
	 	{\it JHEP} {\bf 1202} 026 (2012), arXiv:1109.3230.

\bibitem{HR_bi}
	 	S. F. Hassan, Rachel A. Rosen,
	 	%Bimetric Gravity from Ghost-free Massive Gravity,
{\it JHEP}
{\bf 1202} 126 (2012),	 	 	arXiv:1109.3515.


\bibitem{Kluson}
    J. Kluson,  
%Note About Hamiltonian Structure of Non-Linear Massive Gravity, 
{\it JHEP} {\bf 1201} (2012) 13; 
arXiv:1109.3052.

\bibitem{HR2}
S. F. Hassan, Rachel A. Rosen, 
	%Confirmation of the Secondary Constraint and Absence of Ghost in Massive   Gravity and Bimetric Gravity, 
{\it JHEP} {\bf 1204}  123 (2012),
	arXiv:1111.2070.

\bibitem{Golovnev}
 A. Golovnev,
%    On the Hamiltonian analysis of non-linear massive gravity
{\it Phys. Lett. B} {\bf  707} 404-408 (2012); arXiv:1112.2134.

\bibitem{Krasnov}
S.~Alexandrov, K.~Krasnov, and S.~Speziale.
Chiral description of ghost-free massive gravity;
arXiv:1212.3614; S.~Alexandrov. Canonical structure of Tetrad Bimetric Gravity; arXiv:1308.6586.

\bibitem{ADM}
R.~Arnowitt, S.~Deser and Ch.W.~Misner, in  Gravitation, an Introduction
to Current Research, ed. L.~Witten,  Wiley, New York (1963); arXiv:gr-qc/0405109. 

\bibitem{Kuchar}
 K. Kucha\u{r},  {\it J.~Math.~Phys.} %Vol.
 {\bf 17} 777-791; 792-800; 801-820 (1976);
{\bf 18} 1589-1597 (1977).



\bibitem{SolTch}
V.O. Soloviev, M.V. Tchichikina.
%Bigravity in Kucha\u{r}'s Hamiltonian formalism. 1. The general case. 
{\it Theor. Math. Phys.}, {\bf 176} (3) 1163-1175 (2013); 
arXiv:1211.6530. 

\bibitem{Dirac}
P.A.M. Dirac,  Lectures on Quantum Mechanics. Yeshiva University, New York, (1964). 


\bibitem{Kluson:2012ps}
J.~Kluson. Hamiltonian formalism of particular bimetric gravity model;
arXiv:1211.6267.



\bibitem{Comelli2013}
D. Comelli, F. Nesti and L. Pilo.
Weak Massive Gravity;
 arXiv:1302.4447.

\bibitem{Solo88}
V.O.~Solov'ev, {\it Soviet J. Particles \& Nuclei.} {\bf 19} 482-497 (1988).

\bibitem{Comelli2012}
D. Comelli, M. Crisostomi, F. Nesti and L. Pilo.
%Degrees of freedom in massive gravity
{\it Phys. Rev. D} {\bf 86} 101502(R) (2012);
arXiv:1204.1027.

\bibitem{Leznov}
D. Fairlie, A. Leznov.
%General solutions of the Monge-Amp\`ere equation in $n$-dimensional space.
{\it J. Geom. Phys.} {\bf 16} 385-390 (1995); arxiv:hep-th/9403134.


\bibitem{Deser2}
S. Deser, M. Sandora, A. Waldron.
Nonlinear Partially Massless from Massive Gravity?
	arXiv:1301.5621.

\bibitem{RHRT}
Claudia de Rham, Kurt Hinterbichler, Rachel A. Rosen, Andrew J. Tolley.
Evidence for and Obstructions to Non-Linear Partially Massless Gravity;
 	arXiv:1302.0025.

\bibitem{PMBI1}
S. F. Hassan, Angnis Schmidt-May, Mikael von Strauss.
On Partially Massless Bimetric Gravity;
arXiv:1208.1797.  

\bibitem{PMBI2}
S. F. Hassan, Angnis Schmidt-May, Mikael von Strauss.
Bimetric Theory and Partial Masslessness with Lanczos-Lovelock Terms in Arbitrary Dimensions;
arXiv:1212.4525.

\bibitem{PMBI3}
S. F. Hassan, Angnis Schmidt-May, Mikael von Strauss.
Higher Derivative Gravity and Conformal Gravity From Bimetric and Partially Massless Bimetric Theory
arXiv:1303.6940
 

\bibitem{Rubak}
V.A.~Rubakov and P.G.~Tinyakov,  {\it Phys.-Uspekhi}\  {\bf 51} 759-822 (2008); arXiv:0802.4379.
 
 \bibitem{Blas}
D. Blas. 
 Aspects of Infrared Modifications of Gravity;
arXiv:0809.3744.

\bibitem{MMMM}
 A. Mironov, S. Mironov, A. Morozov and A. Morozov.
Resolving puzzles of massive gravity with and without violation of Lorentz symmetry;
arXiv:0910.5243v1.
Linearized Lorentz-violating gravity and discriminant locus in the moduli space of mass terms;
arXiv:0910.5245v1.

\bibitem{Hinter}
Kurt Hinterbichler,
{\it Rev. Mod. Phys.} {\bf 84} 671–710 (2012); arXiv:1105.3735.
%Theoretical aspects of massive gravity

\bibitem{Kluson:2013}
J.~Kluson. Is bimetric gravity really ghost free?
arXiv:1301.3296.

\bibitem{Kluson:2013-2}
J.~Kluson. Hamiltonian Formalism of General Bimetric Gravity;
arXiv:1303.1652.

\bibitem{SolTch2011}
V.O. Soloviev, M.V. Tchichikina.
Ultralocal energy density in massive gravity;
arXiv:1106.5709. 

\bibitem{Comelli_MayDay}
D.~Comelli, F.~Nesti and L.~Pilo.
Massive gravity: a General Analysis;
arXiv:1305.0236.





\end{thebibliography}
\end{document}